\documentclass[epj,final]{svjour}

\usepackage{graphicx,psfrag}
\usepackage{times}
\usepackage{color}
\usepackage{amsfonts,amssymb,stmaryrd,latexsym,amsmath}

\allowdisplaybreaks[1]

\usepackage{hyperref}

\DeclareMathOperator{\si}{si}
\DeclareMathOperator{\Ci}{ci}
\DeclareMathOperator{\pV}{p.V.}

\begin{document}

\title{Three-body scattering problem in the fixed center approximation: the case of attraction}
\titlerunning{Three-body scattering problem in the fixed center approximation: the case of attraction}

\author{Alexander~E.~Kudryavtsev\inst{1}\thanks{\emph{E-mail: kudryavt@itep.ru}} \and Vakhid~A.~Gani\inst{1,2}\thanks{\emph{E-mail: vagani@mephi.ru}} \and Alexander~I.~Romanov\inst{2}\thanks{\emph{E-mail: einzehl@gmail.com}}}
\authorrunning{A.~Kudryavtsev, V.~Gani, A.~Romanov}

\institute{National Research Center Kurchatov Institute, Institute for Theoretical and Experimental Physics, 117218 Moscow, Russia \and National Research Nuclear University MEPhI (Moscow Engineering Physics Institute), 115409 Moscow, Russia}

\abstract{
We study the scattering of a light particle on a bound pair of heavy particles (e.g., the deuteron) within the fixed center approximation in the case of light-heavy attraction, solving the integral equation for the three-body Green's function both in the coordinate and in the momentum space. The results for the three-body scattering amplitude appear to be ambiguous --- they depend on a single real parameter. This parameter may be fixed by a three-body input, e.g., the three-body scattering length. We also solve the integral equation for the three-body Green function in the momentum space, introducing a finite cut-off. We show that all three approaches are equivalent. We also discuss how our approach to the problem matches with the introduction of three-body contact interaction as done by other authors.
}

\maketitle

\section{Introduction}

The three-body problem in quantum mechanics and quantum field theory has recently been extensively discussed, in particular, in relation to weakly bound states of three particles --- the trimers \cite{kunitski,zinner}. Various approaches giving rise to the three-body problem have been used to study, e.g., some of the highly excited states in charmonium \cite{belle01,LHCb01,bes3,panda01} and bottomonium \cite{drutskoy01} spectra, see also review \cite{brambilla01} and references therein. For example, the resonance $X(3872)$ is often considered as a weakly bound $D\bar D^*$ state \cite{voloshin01,voloshin02,tornqvist01,baru03}. A more long-studied system where one also needs to describe three-particle dynamics is the lightest hadronic atoms, in particular, pionic deuterium \cite{strauch01,hoferichter01,hoferichter02} and kaonic deuterium \cite{mai01}, see also review \cite{gasser01}.

The three-particle dynamics can be treated by the Faddeev equations that give the exact answer for the three-particle scattering amplitude. However, one in general has to know the full two-body $t$-matrix $t(p,p^\prime;E)$ in order to solve the Faddeev equations. Such information is not available for many systems: one has to use approximations. The above examples all share a common feature, namely, one of the particles being light compared to the other two that form a bound state. The dynamics of such systems can be studied within the fixed center approximation (FCA), which treats the two heavy particles as infinitely heavy. This approach has been applied to such systems as $\rho K\bar{K}$, $\eta K\bar{K}$, $\bar{K}NN$ and so on, see, e.g., \cite{bayar01,oset01,oset02,oset03,oset04,oset05}.

Another approximation often used to solve the three-body problem is the Skornyakov-Ter-Martirosyan (STM) equation \cite{STM}. This equation uses only two-body inputs, namely, the two-particle scattering length. It is clear, however, that in order to describe three-particle dynamics such as the (low-energy) scattering phases and the three-particle bound states one needs to complement this information by three-body inputs, e.g., the three-particle scattering length.

The standard approach to solve the STM equation is to introduce ultraviolet cut-off with the cut-off parameter $\Lambda$ in the integral equation in the momentum space~\cite{BHvK}. The dependence of the three-particle scattering length $a_3$ on $\Lambda$ is eliminated by adding a contact three-particle interaction with the $\Lambda$-dependent strength $H(\Lambda)$. The functional form of the latter is chosen such as to make $a_3$ independent of $\Lambda$. However, a new parameter $\Lambda^*$ emerges, such that $a_3=a_3(\Lambda^*)$ and $H=H(\Lambda,\Lambda^*)$.

In this article we discuss the details of the dependence of the three-particle scattering length $a_3$ on $\Lambda$, using an exactly solvable model as an example. This model considers scattering in the three-particle system of a light particle (the $\pi$-meson, having the mass $m_\pi$) interacting with two heavy particles (the nucleons, having the mass $m_N$) that form a bound state (the deuteron). Within the fixed center approximation, we find an exact analytic solution for $a_3$ in this system, and clarify the nature of the cyclic dependence of $a_3$ on $\Lambda$. We also discuss the relation of our solution to the results of \cite{BHvK}.

Our paper is organized as follows. In Section \ref{sec:level2} we write out and solve the FCA equation in the coordinate space. In Section \ref{sec:level3} we solve the FCA equation in the momentum space and demonstrate that the two solutions coincide. In Section \ref{sec:level4} we introduce the cut-off parameter $\Lambda$ and show that the absence of a definite limit of the resulting solution at $\Lambda\to+\infty$ is connected with the ambiguity of the solution of the problem without the cut-off. The parameter $\Lambda$ can be viewed as a physical parameter, and the three-particle scattering length $a_3$ depends on $\Lambda$. We conclude with a discussion of the results in Section~\ref{sec:level5}. Selected technical details are presented in the Appendices.

\section{\label{sec:level2} Fixed Center Approximation in the coordinate space}

The multiple scattering series (MSS) plays an important role in the description of interactions of mesons with nuclei. The expression for the sum of all terms that correspond to the rescattering of a meson on a pair of fixed nucleons was obtained by L.~Foldy in 1945 \cite{Foldy45}. This result was applied to $\pi d$-scattering by K.~Br\"uckner in 1953 \cite{Bruckner531,Bruckner532}. According to Br\"uckner, the $\pi d$-scattering amplitude is the FCA amplitude weighted with the deuteron wave function:
\begin{equation}\label{eq_1}
  F_{\pi d}=\int |\psi_d(\vec{r})|^2 \frac{f_1+f_2+2\frac{f_1f_2}{r}e^{ikr}}{1-\frac{f_1f_2}{r^2}e^{2ikr}}d\vec{r}.
\end{equation}
Here $f_1$ and $f_2$ are the amplitudes of the pion scattering on the first and the second nucleon, respectively, and $\psi_d(\vec{r})$ is the deuteron wave function, normalized by $\int |\psi_d(\vec{r})|^2d\vec{r}=1$. Equation \eqref{eq_1} can also be obtained  by summing up the multiple scattering Feynman graphs \cite{KolKud72}. More recently the MSS terms have been discussed in the context of the effective field theory (EFT) approach to pion-nucleus scattering, starting from the first EFT calculation of $\pi d$ scattering made by Weinberg \cite{Wein92}. Various aspects of MSS related to the EFT formalism have been studied in the next twenty years, see, e.g., \cite{beane01,beane02,baru01,lensky01,baru02}. In particular, Ref.~\cite{Baru} showed in the framework of the EFT that the divergences of individual MSS terms cancel each other.

Here we work with the expression given by \eqref{eq_1}, concentrating on the case of pion-nucleon attraction. We assume $f_1=f_2=a$. This results in the following expression for the scattering amplitude at zero energy (i.e., the scattering length):
\begin{equation}
a_3=F^{(S)}(a)+F^{(M)}(a),
\end{equation}
where $F^{(S)}(a)=2\bar{a}/(1+\xi/2)$ is the single scattering contribution, $\bar{a}=a(1+\xi)$, $\xi=m_{\pi}/m_N$, and $F^{(M)}(a)$ is the sum of all multiple scattering terms. Note that only the rescaled scattering length $\bar{a}=a(1+\xi)$ enters the expressions from this point on, and we omit the bar for convenience. The coordinate space expression for $F^{(M)}(a)$ reads
\begin{equation}\label{eqn3}
F^{(M)}(a)=\frac{2a^2}{1+\xi/2}\int\frac{|\psi_d(\vec{r})|^2}{r-a}d\vec r.
\end{equation}
In the case of pion-nucleon attraction, $a>0$, the integral in Eq.~\eqref{eqn3} is divergent. As we will demonstrate, the integration kernel $\displaystyle\frac{1}{r-a}$ in this case should be replaced by the operator
\begin{equation}\label{eq_4}
\hat A(r)=\pV \frac{1}{r-a}+\mathbb{B}\cdot\delta (r-a),
\end{equation}
where p.V.\ stands for the principal value of the integral, and $\mathbb{B}$ is a dimensionless parameter undetermined by the equations that needs to be fixed from experiment. This gives
\begin{equation}\label{eq_5}
F^{(M)}(a)=\frac{2a^2}{1+\xi/2}\int\hat A(r)\:|\psi_d(\vec{r})|^2\:d\vec{r}.
\end{equation}
We conclude that for a given $a>0$ the three-particle scattering length $a_3$ is ambiguous; it depends on the arbitrary constant $\mathbb{B}$. Knowing $a_3$ one can determine $\mathbb{B}$ and in this way predict the energy dependence of the three-particle $s$-wave scattering phase.

\section{\label{sec:level3} Fixed Center Approximation in the momentum space}

Considering the Feynman graphs it is easy to obtain the $n$-tuple rescattering amplitudes $f^{(n)}(a)$ within the FCA (see, e.g., \cite{Bruckner531,Bruckner532}):
\begin{equation}\label{2_1}
\begin{split}
f^{(n)}(a)=\int\frac{\varphi_d(\vec{p})}{(2\pi)^3}&\Sigma^{(n)}(\vec{p},\vec{p}^{\phantom{i}\prime})\frac{\varphi_d(\vec{p}^{\phantom{i}\prime})}{(2\pi)^3}d\vec{p}d\vec{p}^{\phantom{i}\prime},\\
&n=2,3,...,
\end{split}
\end{equation}
where $\varphi_d(\vec{p})$ is the deuteron wave function in the momentum space, normalized by $\int |\varphi_d(\vec{p})|^2d\vec{p}=(2\pi)^3$, and
\begin{equation}\label{2_2}
\begin{split}
\Sigma^{(2)}(\vec{p},\vec{p}^{\phantom{i}\prime})=\frac{2a^2}{1+\xi/2}&\frac{4\pi}{(\vec{p}-\vec{p}^{\phantom{i}\prime})^2},{}\\
\Sigma^{(3)}(\vec{p},\vec{p}^{\phantom{i}\prime})=\frac{2a^3}{1+\xi/2}&\int\frac{d\vec{s}}{(2\pi)^3}
\frac{4\pi}{(\vec{p}-\vec{s})^2}\frac{4\pi}{(\vec{s}-\vec{p}^{\phantom{i}\prime})^2},{}\\
\Sigma^{(4)}(\vec{p},\vec{p}^{\phantom{i}\prime})=\frac{2a^4}{1+\xi/2}&\int\frac{d\vec{s}d\vec{t}}{(2\pi)^6}
\frac{4\pi}{(\vec{p}-\vec{s})^2}\frac{4\pi}{(\vec{s}-\vec{t})^2}
\frac{4\pi}{(\vec{t}-\vec{p}^{\phantom{i}\prime})^2},
\end{split}
\end{equation}
and so on. Taking into account only the leading $s$-wave part of the deuteron wave function, we can integrate over the angles in \eqref{2_1}, obtaining
\begin{equation}\label{2_3}
\begin{split}
&f^{(2)}(a)=\frac{2a^2}{1+\xi/2}\frac{(4\pi)^2}{(2\pi)^6}\times\\
&\times\int\limits_0^{+\infty}p\varphi_d(p)\cdot\pi\ln(p,p^\prime)\cdot\varphi_d(p^\prime)p^\prime dpdp^\prime,\\
&f^{(3)}(a)=\frac{2a^3}{1+\xi/2}\frac{(4\pi)^2}{(2\pi)^6}\int\limits_0^{+\infty}p\varphi_d(p)\times\\
&\times\left(\int\limits_0^{+\infty}\pi\ln(p,s)\cdot\pi\ln(s,p^\prime)\frac{ds}{2\pi^2}\right)\varphi_d(p^\prime)p^\prime dpdp^\prime,
\end{split}
\end{equation}
and so on, where
\begin{equation}
\ln(p,p^\prime)=\ln\left(\frac{p+p^\prime}{p-p^\prime}\right)^2.
\end{equation}
The full multi-scattering amplitude $F^{(M)}(a)=\sum\limits_{n=2}^{+\infty}f^{(n)}(a)$ in the momentum space is (see Ref.~\cite{KRG})
\begin{equation}\label{2_4}
\begin{split}
F^{(M)}(a)&=\frac{2a^2}{1+\xi/2}\frac{(4\pi)^2}{(2\pi)^6}\times\\
&\times\int\limits_0^{+\infty}p\varphi_d(p)\cdot R(p,p^\prime)\cdot\varphi_d(p^\prime)p^\prime dpdp^\prime,
\end{split}
\end{equation}
where the function $R(p,p^\prime)$ is the solution of the following integral equation:
\begin{equation}\label{eq_11}
R(p,p^\prime)=\pi\ln(p,p^\prime)+\frac{a}{2\pi^2}\int\limits_0^{+\infty}\pi\ln(p,s)R(s,p^\prime)ds.
\end{equation}

First, we consider the integral equation \eqref{eq_11} with the infinite upper limit of integration. The corresponding solution is given by the following integral (see the Appendix A for the derivation):
\begin{equation}\label{eq_12}
\begin{split}
&R_\infty(p,p^\prime)=\\
&4\pi\int\limits_0^{+\infty}\sin pr\sin p^\prime r\left(\pV\frac{1}{r-a}+\mathbb{B}\cdot\delta(r-a)\right)dr.
\end{split}
\end{equation}
The amplitude $F^{(M)}(a)$ is a matrix element of the operator $R(p,p^\prime)$, see Eq.~\eqref{2_4}. From Eq.~\eqref{eq_12} we see that $R_\infty(p,p^\prime)$ can be represented as a sum:
\begin{equation}\label{eq:Rinf}
R_\infty(p,p^\prime)=R_\mathrm{In}(p,p^\prime)+R_\mathrm{Hom}(p,p^\prime),
\end{equation}
where the function
\begin{equation}\label{eqn14}
R_\mathrm{In}(p,p^\prime)=4\pi \pV\int\limits_0^{+\infty} \frac{\sin pr\sin p^\prime r}{r-a}dr
\end{equation}
is a solution of the inhomogeneous equation \eqref{eq_11} as can be checked by substitution. As shown in the Appendix B, the function $R_\mathrm{In}(p,p^\prime)$ takes the following form for $p>p^\prime$:
$$
R_\mathrm{In}(p,p^\prime)=2\pi\left[\cos(aP_+)\cdot\Ci(aP_+)-\sin(aP_+)\cdot\si(-aP_+)\right]
$$
\begin{equation}\label{eq_14}  -2\pi\left[\cos(aP_-)\cdot\Ci(aP_-)-\sin(aP_-)\cdot\si(-aP_-)\right],
\end{equation}
and for $p'>p$:
$$R_\mathrm{In}(p,p^\prime)=2\pi\left[\cos(aP_+)\cdot\Ci(aP_+)-\sin(aP_+)\cdot\si(-aP_+)\right]
$$
\begin{equation}\label{eqn16}
-2\pi\left[\cos(a|P_-|)\cdot\Ci(aP_-)+\sin(aP_-)\cdot\si(-a|P_-|)\right],
\end{equation}
where $P_+=p+p^\prime$, $P_-=p-p^\prime$, with $\Ci(x)$ and $\si(x)$ being the integral cosine and the integral sine, respectively \cite{JEL}. It follows from Eqs.~\eqref{eq_14} and \eqref{eqn16} that $R_\mathrm{In}(p,p^\prime)=R_\mathrm{In}(p^\prime,p)$. The asymptotic expression for $R_\mathrm{In}(p,p^\prime)$ at $p\gg p^\prime$ can be obtained from Eq.~\eqref{eq_14} (see the Appendix B) and reads
\begin{equation}\label{eq_16}
R_\mathrm{In}^\mathrm{(as)}(p,p^\prime)=4\pi^2\cos pa\sin p^\prime a.
\end{equation}
The general solution of the homogeneous equation, $R_\mathrm{Hom}(p,p^\prime)$, is
\begin{equation}\label{eq_17}
R_\mathrm{Hom}(p,p^\prime)=4\pi\mathbb{B}\sin ap\sin ap^\prime.
\end{equation}
The integral equation \eqref{eq_11} in the momentum space yields the same solution as obtained above in the coordinate space. Below we compare these results with those obtained by solving the integral equation introducing a finite cut-off $\Lambda$, and discuss the $\Lambda$-dependence of the solution.

\section{\label{sec:level4} The problem of $\Lambda$-dependence in the momentum space}

The function $R_\infty(p,p^\prime)$, defined by \eqref{eq_12}, is a solution of the integral equation \eqref{eq_11}. This equation could be solved (e.g., numerically) by replacing the infinite upper integration limit by a finite cut-off $\Lambda$. References \cite{Baru,KRG} showed that the function $R_{\Lambda}(p,p^\prime)$ obtained in that way strongly depends on the value of $\Lambda$, see also \cite{Dan,Khar}. There seems thus to be no way to get a solution of Eq.~\eqref{eq_11} that would not depend on $\Lambda$ asymptotically in the limit of large $\Lambda$. We will try to solve this problem.

First of all, we have to assume that the solutions of Eq.~\eqref{eq_11} with and without the cut-off coincide at large $\Lambda$, at least asymptotically. It means that the solutions $R_\mathrm{In}(p,p^\prime)$ and $R_\mathrm{Hom}(p,p^\prime)$, see Eqs.~\eqref{eq_14} and \eqref{eq_17}, fulfill the equation with the cut-off at large $\Lambda$.

From the analysis of the numerical solution of Eq.~\eqref{eq_11} with the cut-off one can see \cite{private} that in the limit $p\gg p^\prime$, $p\lesssim\Lambda$ the function $R_{\Lambda}(p,p^\prime)$ takes the form
\begin{equation}\label{eq_20}
R_{\Lambda}(p,p^\prime)=A(\Lambda)\cdot\sin(a(p-\Lambda)+\varphi_0)\cdot\sin p^\prime a,
\end{equation}
where $\varphi_0$ does not depend on $\Lambda$. Let us show that the sum of the functions $R_\mathrm{In}(p,p^\prime)$ and $R_\mathrm{Hom}(p,p^\prime)$ at $p\gg p^\prime$, $p\lesssim\Lambda$ is consistent with the asymptotic expression \eqref{eq_20}. The asymptotic of this sum is
\begin{equation}\label{eqn22}
R_{\infty}^\mathrm{(as)}(p,p^\prime)=4\pi^2(\cos pa+b\sin pa)\cdot\sin p^\prime a,
\end{equation}
where $b=\mathbb{B}/\pi$. It is convenient to rewrite \eqref{eqn22} in the form
\begin{equation}
R_{\infty}^\mathrm{(as)}(p,p^\prime)=\frac{4\pi^2}{\sin\phi}\sin(pa+\phi)\cdot\sin p^\prime a,
\end{equation}
where $\sin\phi=1/\sqrt{1+b^2}$ and $\cos\phi=b/\sqrt{1+b^2}$. Notice that $R_{\infty}^\mathrm{(as)}(p,p^\prime)$ coincides with $R_{\Lambda}(p,p^\prime)$ \eqref{eq_20} if $\phi=a(\Lambda_\mathrm{cr}^{(i)}-\Lambda)$ and $A(\Lambda)=4\pi^2/\sin a(\Lambda_\mathrm{cr}^{(i)}-\Lambda)$. One thus gets
\begin{equation}\label{eqn23}
\begin{split}
&R_{\Lambda}^\mathrm{(as)}(p,p^\prime)=\\
&=\frac{4\pi^2}{\sin a(\Lambda_\mathrm{cr}^{(i)}-\Lambda)}\cdot\sin[(p+\Lambda_\mathrm{cr}^{(i)}-\Lambda)a]\sin p^\prime a.
\end{split}
\end{equation}
This expression is only valid in the limit $p\gg p^\prime$, $p\lesssim\Lambda$. The parameter $\Lambda_\mathrm{cr}^{(i)}$ in \eqref{eqn23} is the critical point nearest to $\Lambda$, i.e.~the point where $R_{\Lambda}(p,p^\prime)$ goes to infinity, see, e.g., \cite{KRG}. Any two adjacent critical points $\Lambda_\mathrm{cr}^{(i)}$ are separated by approximately the same distance on the $\Lambda$ axis \cite{private}, i.e.\
\begin{equation}
\Lambda_\mathrm{cr}^{(i)}=\Lambda_\mathrm{cr}^{(1)}+\Delta\Lambda\cdot(i-1).
\end{equation}
$\Lambda_\mathrm{cr}^{(i)}$ in Eq.~\eqref{eqn23} can be replaced by any $\Lambda_\mathrm{cr}^{(j)}$, and, in particular, by $\Lambda_\mathrm{cr}^{(1)}$.

So we obtained the general solution for $R_{\Lambda}(p,p^\prime)$ in the asymptotics $p\gg p^\prime$, $p\lesssim\Lambda$:
\begin{equation}\label{eq_22}
R^\mathrm{sum}_{\Lambda}(p,p^\prime)=R_\mathrm{In}(p,p^\prime)+R_\mathrm{Hom}(p,p^\prime),
\end{equation}
where $b=\mathbb{B}/\pi$ is connected with the cut-off $\Lambda$ through the parameter $\phi$, see above. As we shall see later from the numerical analysis, this expression is valid for all $p,p^\prime\lesssim\Lambda$, not only at $p\gg p^\prime$.

We thus arrive at a conclusion that the solution $R_{\Lambda}(p,p^\prime)$ can at arbitrary $\Lambda$ be expressed through the sum of the solutions of the homogeneous and inhomogeneous main equation \eqref{eq_11} with infinite upper integration limit. This means that the solutions of Eq.~\eqref{eq_11} with the cut-off $\Lambda$ do not bear any new information. One can, in fact, eliminate the parameter $\Lambda$ by substituting
\begin{equation}\label{eq_23}
\cot a(\Lambda_\mathrm{cr}^{(i)}-\Lambda)=b.
\end{equation}
where the parameter $b$ is chosen to reproduce the empirical value of $a_3$, and the nearest critical value $\Lambda_\mathrm{cr}^{(i)}>\Lambda$ should be used. This connection between $\Lambda$ and $b$ leads to a good agreement between the solutions for all $p,p^\prime\lesssim\Lambda$, as we shall see later.

We solved the integral equation \eqref{eq_11} with the cut-off numerically, using a rectangular grid in the $(p,s)$ plane, while $p^\prime$ was fixed to $p^\prime=50$ MeV. The grid spacing was adjusted in order to achieve the desired accuracy (we used the grid spacing 5 MeV).

\begin{figure}
\centering
\includegraphics[scale=0.75]{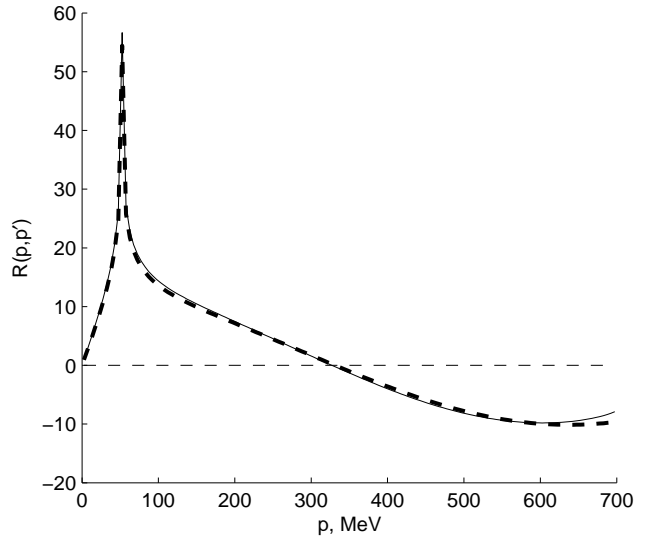}
\caption{Solution of the integral equation \eqref{eq_11} with the cut-off $\Lambda=700$ MeV (solid line) and without the cut-off for $b=0.07$ (dashed line, $\Lambda_\mathrm{cr}^{(2)}=1000$ MeV). Here $p^\prime=50$ MeV, $a=0.005$ MeV$^{-1}$.}
\label{Fig1}
\end{figure}

\begin{figure}
\centering
\includegraphics[scale=0.75]{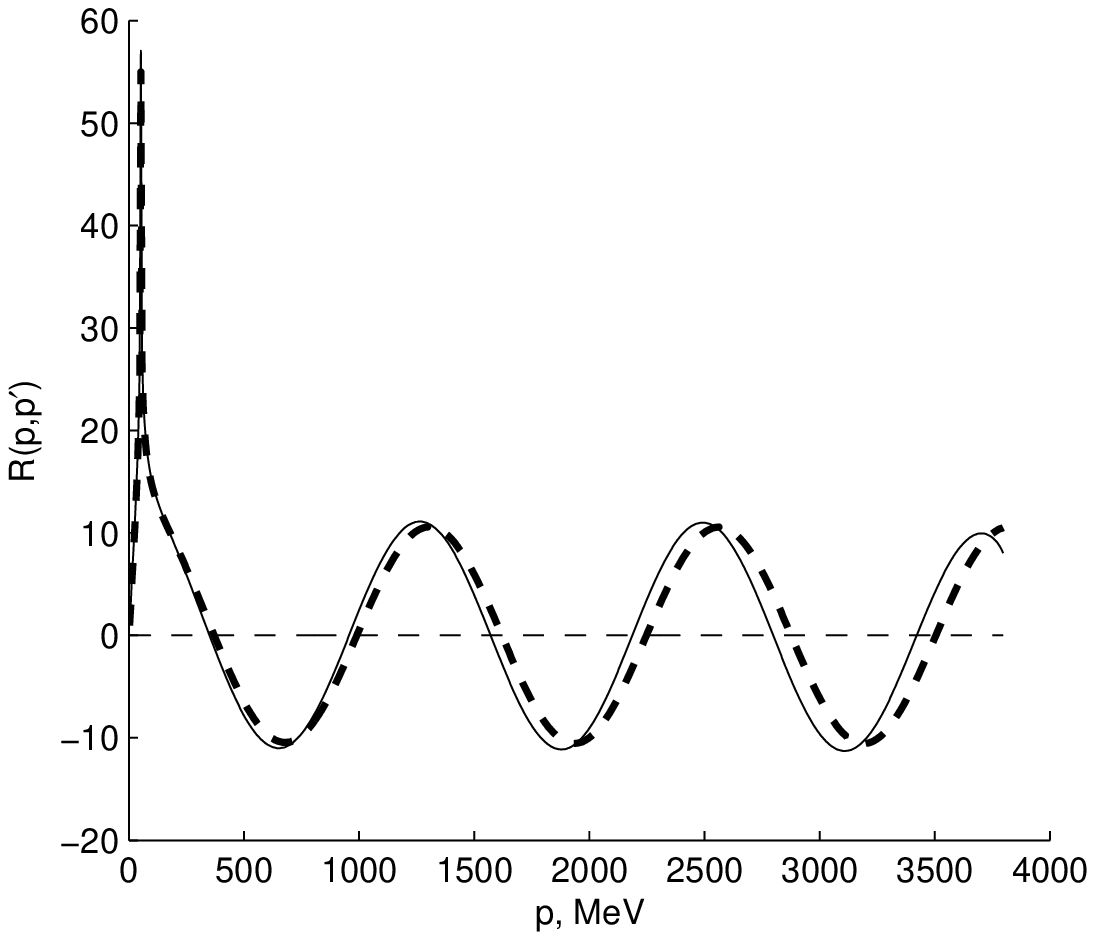}
\caption{Solution of the integral equation \eqref{eq_11} with the cut-off $\Lambda=3800$ MeV (solid line) and without the cut-off for $b=0.20$ (dashed line, $\Lambda_\mathrm{cr}^{(7)}=4075$ MeV). Here $p^\prime=50$ MeV, $a=0.005$ MeV$^{-1}$.}
\label{Fig2}
\end{figure}

\begin{figure}
\centering
\includegraphics[scale=0.75]{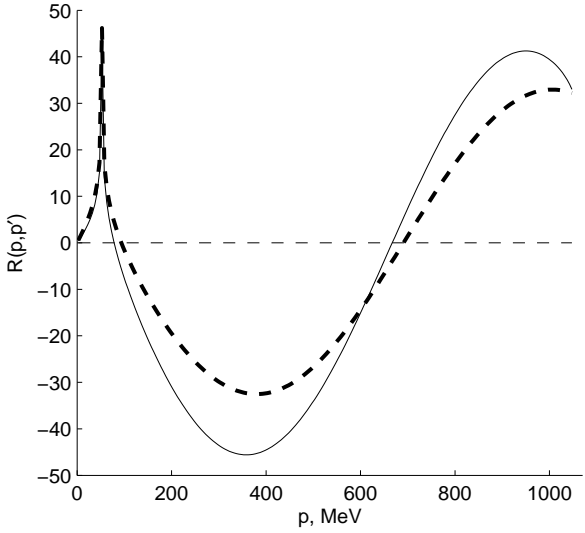}
\caption{Solution of the integral equation \eqref{eq_11} with the cut-off $\Lambda=1050$ MeV (solid line) and without the cut-off for $b=-3.05$ (dashed line, $\Lambda_\mathrm{cr}=1615$ MeV). Here $p^\prime=50$ MeV, $a=0.005$ MeV$^{-1}$.}
\label{Fig3}
\end{figure}

In Figs.~\ref{Fig1} -- \ref{Fig3} we compare the solutions of Eq.~\eqref{eq_11} obtained with and without the cut-off. Our choice of the upper integration limit $\Lambda$ for the numerical solution is dictated by the positions of the critical values $\Lambda_\mathrm{cr}^{(i)}$. In particular, $a=0.005$ MeV$^{-1}$ corresponds to $\Lambda_\mathrm{cr}^{(1)}=385$ MeV, $\Lambda_\mathrm{cr}^{(2)}=1000$ MeV, $\Lambda_\mathrm{cr}^{(3)}=1615$ MeV, and so on.

In Fig.~\ref{Fig1} we show the function $R_{\Lambda}(p,p^\prime)$ for $p^\prime=50$ MeV and $\Lambda=700$ MeV. The latter value is roughly halfway between the first and the second critical values of $\Lambda$. The corresponding parameter $b$, which is given by Eq.~\eqref{eq_23}, appears to be small ($b=0.07$).

Figure~\ref{Fig2} demonstrates the oscillating character of the asymptotic of the solution. In order to do so, we selected a rather large value of the cut-off, $\Lambda=3800$ MeV, which is also roughly at the midpoint between two adjacent critical values of $\Lambda$.

Furthermore, Fig.~\ref{Fig3} shows the results corresponding to $\Lambda=1050$ MeV, which is close to $\Lambda_\mathrm{cr}^{(2)}$. In this case the absolute value of the parameter $b$ appears to be large ($b=-3.05$).

Figures \ref{Fig1} and \ref{Fig2} show that the numerical solution of the integral equation (solid curve) almost coincides with its analytical solution (dashed curve) if $\Lambda$ is far from critical values. However, the agreement becomes worse if $\Lambda$ is close to a critical value, see Fig.~\ref{Fig3}. This may be a consequence of the finite accuracy of the numerical extraction of $\Lambda_\mathrm{cr}^{(i)}$.

\section{\label{sec:level5} Conclusion}

We have studied the process of scattering of a light particle on a pair of heavy particles within the fixed center approximation. We have shown that the summation of the multiple scattering series gives the same results both in the coordinate and in the momentum representation. In the momentum representation the general solution $R(p,p^\prime)$ of the integral equation \eqref{eq_11} for the Green's function can be expressed as a sum of the general solution of the corresponding homogeneous equation and a partial solution of the inhomogeneous equation. The solution of the homogeneous equation is defined up to a constant factor. We have obtained analytical expressions for a partial solution of the inhomogeneous equation and for the general solution of the corresponding homogeneous equation. We have also obtained an analytical expression for the solution of the inhomogeneous equation at $p\gg p^\prime$ \eqref{eq_16}.

We have solved the integral equation \eqref{eq_11} by introducing a finite integration cut-off $\Lambda$. We analyzed the corresponding numerical solutions in the limit $p\gg p^\prime$, $p\lesssim\Lambda$ and identified the dependence \eqref{eq_20} of the solution on the cut-off. This dependence being periodic has been argued to be a consequence of the ambiguity of the solution of Eq.~\eqref{eq_11} without the cut-off. Within this approach, we have obtained a relation between the cut-off $\Lambda$ and the constant $b$ (or $\mathbb{B}$) which is incorporated in the solution of the homogeneous equation without the cut-off. Apart from that, in the numerical calculations we observed that the correspondence between the solutions with and without the cut-off holds for all $p,p^\prime\lesssim\Lambda$, i.e., not only in the asymptotic limit $p\gg p^\prime$.

The obtained results can be interpreted in the following way. The periodic dependence of the solution \eqref{eq_20} on the cut-off $\Lambda$ (i.e., the absence of a limit at $\Lambda\to+\infty$) is connected to the ambiguity of the solution of Eq.~\eqref{eq_11} with the infinite upper limit of integration. This ambiguity does not allow one to predict the three-body scattering length $a_3$. On the other hand, the single parameter that each of the solutions depends on --- $\mathbb{B}$ (or $b$) without the cut-off or $\Lambda$ with the cut-off --- can be fitted to the empirical value of $a_3$. After adjusting the parameter, the theory provides a unique answer. The procedure proposed by us is in that sense different from that used in, e.g.,~\cite{BHvK}, where a three-body contact interaction is introduced in order to eliminate the dependence of $a_3$ on $\Lambda$. 

We would also like to briefly comment on the STM equation. Unlike the STM, the FCA equation is exactly solvable, however, the solutions of the two equations have much in common. The solution of the STM equation with the infinite upper limit of integration is ambiguous because of the solution of the corresponding homogeneous equation. After introducing the cut-off $\Lambda$, the same cyclic dependence of the answer on $\Lambda$ appears in the STM equation, as in the FCA. So we can assume that the solutions of the STM with the cut-off can be expressed through the solutions of the STM without the cut-off, and after that, the free parameter (the cut-off $\Lambda$ or another constant if the equation is solved with the infinite upper integration limit) should be fixed in order for the theory to reproduce the empirical value of the three-body scattering length $a_3$.

\section*{Acknowledgments}

The authors would like to thank J.~Gegelia and C.~Hanhart for very useful discussions of the main results of this work, K.~G.~Boreskov, L.~N.~Bogdanova, and V.~G.~Ksenzov for useful discussion during the theory seminar at ITEP, V.~V.~Baru and E.~Epelbaum for constructive criticism of the first version of the manuscript and for useful remarks. The authors are also very grateful to V.~Lensky for critical comments that resulted in substantial improvement of the manuscript.

This work is supported in part by the DFG and the NSFC through funds provided to the Sino-German CRC 110 ``Symmetries and the Emergence of Structure in QCD'' (NSFC Grant No. 11261130311). This research was also partially supported by the MEPhI Academic Excellence Project (contract number 02.03.21.0005, 27.08.2013).

\appendix
\renewcommand{\theequation}{\Alph{section}.\arabic{equation}}
\setcounter{equation}{0}

\section{Solution of the main integral equation \eqref{eq_11} with the infinite upper integration limit}

Let us check that the function \eqref{eq_12}
\begin{equation*}
\begin{split}
&R_\infty(p,p^\prime)=\\
&=4\pi\int\limits_0^{+\infty}\sin(pr)\sin(p^\prime r)\left(\pV\frac{1}{r-a}+\mathbb{B}\cdot\delta(r-a)\right)dr
\end{split}
\end{equation*}
is a solution of Eq.~\eqref{eq_11}:
$$
R(p,p^\prime)=\pi\ln(p,p^\prime)+\frac{a}{2\pi^2}\int\limits_0^{+\infty}\pi\ln(p,s)R(s,p^\prime)\:ds.
$$
Plugging in the first term of the function $R_{\infty}(p,p^\prime)$ into the integral in Eq.~\eqref{eq_11}, we get
$$
\int\limits_0^{+\infty}\pi\:\ln\left(\frac{p+s}{p-s}\right)^2ds\cdot 4\pi\pV\int\limits_0^{+\infty}\frac{\sin(sr)\sin(p^\prime r)}{r-a}\:dr=
$$
$$
=4\pi^2\int\limits_0^{+\infty}ds\cdot 4\int\limits_0^{+\infty}\frac{\sin(pr^\prime)\sin(sr^\prime)}{r^\prime}\:dr^\prime\times
$$
$$\times\pV\int\limits\frac{\sin(sr)\sin(p^\prime r)}{r-a}\:dr=
$$
\begin{equation}
=8\pi^3\cdot\pV\int\limits_0^{+\infty}\frac{\sin(pr)\sin(p^\prime r)}{r(r-a)}\:dr.
\end{equation}
Here we used the following identities:
\begin{equation}
\ln\left(\frac{p+s}{p-s}\right)^2=4\int\limits_0^{+\infty}\frac{\sin(pr)\sin(sr)}{r}\:dr,
\label{eqLog}
\end{equation}
$$
2\int\limits_0^{+\infty}\sin(sr^\prime)\sin(sr)ds=
$$
$$
=\int\limits_0^{+\infty}\left[\cos(s(r-r^\prime))-\cos(s(r+r^\prime))\right]ds=
$$
\begin{equation}
=\pi\left[\delta(r-r^\prime)-\delta(r+r^\prime)\right].
\end{equation}
Then the l.h.s.\ and the r.h.s.\ of the integral equation \eqref{eq_11} yield:
$$
4\pi \pV\int\limits_0^{+\infty}\frac{\sin(pr)\sin(p^\prime r)}{r-a}\:dr
=4\pi\int\limits_0^{+\infty}\frac{\sin(pr)\sin(p^\prime r)}{r}\:dr
\:+
$$
\begin{equation}
+\:4\pi a \pV\int\limits_0^{+\infty}\frac{\sin(pr)\sin(p^\prime r)}{r(r-a)}\:dr,
\end{equation}
or
\begin{equation}
\frac{1}{r-a}=\frac{1}{r}+\frac{a}{r(r-a)}.
\end{equation}
This confirms that \eqref{eq_12} is a solution of Eq.~\eqref{eq_11}.

\section{The particular solution of the inhomogeneous equation \eqref{eq_11}, and its asymptotics}

\setcounter{equation}{0}

Equation \eqref{eqn14} for $R_\mathrm{In}(p,p^\prime)$ can be rewritten with the help
of the integral sine and integral cosine. Substituting $\sin pr\sin p^\prime r$ by $\displaystyle\frac{1}{2}\left[\cos(P_-r)-\cos(P_+r)\right]$, where $P_+=p+p^\prime$, $P_-=p-p^\prime$, we obtain:
$$
4\pi\pV\int\limits_0^{+\infty}\frac{\sin(pr)\sin(p^\prime r)}{r-a}\:dr=
$$
$$
=2\pi\left[\pV\int\limits_0^{+\infty}\frac{\cos\left[P_-(r-a)\right]}{r-a}dr\cdot\cos(P_-a)-\right.
$$
$$
\left.-\pV\int\limits_0^{+\infty}\frac{\sin\left[P_-(r-a)\right]}{r-a}dr\cdot\sin(P_-a)\right]-
$$
$$
-2\pi\left[\pV\int\limits_0^{+\infty}\frac{\cos\left[P_+(r-a)\right]}{r-a}dr\cdot\cos(P_+a)-\right.
$$
\begin{equation}
\left.-\pV\int\limits_0^{+\infty}\frac{\sin\left[P_+(r-a)\right]}{r-a}dr\cdot\sin(P_+a)\right].
\end{equation}
Taking into account that
\begin{equation}
\pV\displaystyle\int\limits_0^{+\infty}\frac{\cos\left[P(r-a)\right]}{r-a}dr=-\Ci(-aP),
\end{equation}
and
\begin{equation}
\pV\displaystyle\int\limits_0^{+\infty}\frac{\sin\left[P(r-a)\right]}{r-a}dr=-\si(-aP),
\end{equation}
we finally get for $P_->0$:
$$
R(p,p^\prime)=2\pi\{\Ci(aP_+)\cos(aP_+)-\si(-aP_+)\sin(aP_+)\}-
$$
\begin{equation}
-2\pi\{\Ci(aP_-)\cos(aP_-)-\si(-aP_-)\sin(aP_-)\}.
\end{equation}
In case of $P_-<0$ the second term in the right-hand side of the previous equation
is replaced by
\begin{equation}
-2\pi\{\Ci(-aP_-)\cos(aP_-)+\si(-a|P_-|)\sin(aP_-)\}.
\end{equation}
Using the identity $-\si(-x)=\si(x)+\pi$, we finally get for $P_->0$:
$$
R(p,p^\prime)=2\pi\{\Ci(aP_+)\cos(aP_+)+[\si(aP_+)+\pi]\sin(aP_+)\}-
$$
\begin{equation}
-2\pi\{\Ci(aP_-)\cos(aP_-)+[\si(aP_-)+\pi]\sin(aP_-)\}.
\end{equation}
In the case of $P_-<0$ the second term in the right-hand side has the form
\begin{equation}
-2\pi\{\Ci(-aP_-)\cos(aP_-)-[\si(a|P_-|)+\pi]\sin(aP_-)\}.
\end{equation}
This yields the correct answer for $R_\mathrm{In}(p,p^\prime)$. To obtain the asymptotics at $p\gg p^\prime$ we rewrite $R(p,p^\prime)$ in the following form:
$$
R(p,p^\prime)=2\pi\{\Ci(aP_+)\cos(aP_+)-\Ci(aP_-)\cos(aP_-)\}+
$$
\begin{equation}
+2\pi\{\sin(aP_+)[\pi+\si(aP_+)]-\sin(aP_-)[\pi+\si(aP_-)]\}.
\end{equation}
Here we again used the identity $-\si(-x)=\si(x)+\pi$. The term in the first curly braces vanishes at $p\gg p^\prime$. The term in the second curly braces turns into
\begin{equation}
\begin{split}
R_{In}^{(as)}(p,p^\prime)&=2\pi^2\cdot(\sin(aP_+)-\sin(aP_-))=\\
&=4\pi^2\cos pa\sin p^\prime a.
\end{split}
\end{equation}

\end{document}